\documentclass{Interspeech2024}

\usepackage{multirow}
\usepackage{graphicx}

\interspeechcameraready 

\title{Single-Codec: Single-Codebook Speech Codec towards High-Performance Speech Generation}

\name{Hanzhao Li$^1$,
    Liumeng Xue$^2$,
    Haohan Guo$^3$,
    Xinfa Zhu$^1$,
    Yuanjun Lv$^1$,
    Lei Xie$^{1,*}$, \\
    Yunlin Chen$^4$,
    Hao Yin$^4$,
    Zhifei Li$^4$
}

\address{
  $^1$Audio, Speech and Language Processing Group (ASLP@NPU), School of Computer Science, \\ Northwestern Polytechnical University, Xi'an, China\\
  $^2$School of Data Science, The Chinese University of Hong Kong, \\ Shenzhen (CUHK-Shenzhen), China \\
  $^3$The Chinese University of Hong Kong, Hong Kong SAR, China \\
  $^4$Shanghai Mobvoi Information Technology Co., Ltd}
\email{lihanzhao@mail.nwpu.edu.cn, lxie@nwpu.edu.cn {\thanks{* Corresponding author.}}}


\begin{document}

\maketitle

\begin{abstract}
    

    The multi-codebook speech codec enables the application of large language models (LLM) in TTS but bottlenecks efficiency and robustness due to multi-sequence prediction. To avoid this obstacle, we propose Single-Codec, a single-codebook single-sequence codec, which employs a disentangled VQ-VAE to decouple speech into a time-invariant embedding and a phonetically-rich discrete sequence. Furthermore, the encoder is enhanced with 1) contextual modeling with a BLSTM module to exploit the temporal information, 2) a hybrid sampling module to alleviate distortion from upsampling and downsampling, and 3) a resampling module to encourage discrete units to carry more phonetic information. Compared with multi-codebook codecs, e.g., EnCodec and TiCodec, Single-Codec demonstrates higher reconstruction quality with a lower bandwidth of only 304bps. The effectiveness of Single-Code is further validated by LLM-TTS experiments, showing improved naturalness and intelligibility.

\end{abstract}

\vspace{-1em}

\section{Introduction}

Large language models (LLMs) have attracted wide attention in the speech domain, particularly in text-to-speech synthesis (TTS) ~\cite{Wang2023NeuralCL, Kharitonov2023SpeakRA}. In such LLM-based TTS systems, to operate speech synthesis as a simple next-token prediction problem, the first thing is to seek an appropriate speech codec for speech tokenization and waveform reconstruction. Multi-codebook codecs~\cite{Defossez2022HighFN}, as the SOTA approaches, are widely adopted in LLM-based TTS to achieve superior reconstruction quality. However, they also require the LLM to predict multiple discrete sequences, affecting efficiency and stability seriously, although various designs of codec~\cite{zhang2023speechtokenizer, ji2024language, Ren2023FewertokenNS} and LLM~\cite{Borsos2023SoundStormEP, DBLP:journals/corr/abs-2310-00704, Ziv2024MaskedAG} are proposed to adapt this multi-sequence discrete representation better. Hence, seeking an effective approach to obtain the single-sequence discrete speech representation is critical to bypass this limitation.

However, unlike the text, it is impossible to completely represent the speech audio with abundant information in semantics and acoustics with only one discrete token sequence. Although Tortoise-TTS~\cite{Betker2023BetterSS} achieves the LLM with the single-sequence discrete speech representation, A diffusion model still needs to be trained to generate Mel Spectrograms from latent embeddings in the LLM of predicted speech units. These embeddings contain more information related to the input text and the target speaker to compensate for the compression loss. However, this operation introduces more training and inference costs. Recently, TiCodec~\cite{Ren2023FewertokenNS} proposes introducing an additional global encoder to disentangle time-invariant information out of speech units, reducing the amount of frame-level information that needs encoding. It inspires us to re-think speech codec from the perspective of feature disentanglement.

In this study, we propose a single-codebook neural audio codec, \textit{Single-Codec}, for high-performance speech generation. Single-Codec performs compression and reconstruction on Mel Spectrogram instead of the raw waveform, enabling efficient compression of speech information while preserving important details, as stated in Tortoise-TTS~\cite{Betker2023BetterSS}. To further enhance the codec performance and applicability to speech synthesis, Single-Codec incorporates several key components:
\begin{itemize}
    \item A global reference encoder to decouple time-invariant features. Specifically, we utilize continuous global representations rather than discrete representations and longer reference segments to capture more acoustic details, enabling embedding sufficient phonetic information into single-codebook discrete units.
    \item A BLSTM module for contextual modeling to help discover the correlation between adjacent frames, enhancing speech content clustering efficiency.
    \item A hybrid sampling module that uses both convolution and pooling to achieve downsampling, and transposed convolution and replication to achieve upsampling, alleviating upsampling and downsampling distortion.
    \item A resampling module to encourage the encoder to extract more phonetics-relevant information with lower short-time variance from the acoustic sequence. 
\end{itemize}
To the best of our knowledge, Single-Codec is the first single-codebook codec dedicatedly designed for LLM-based speech generation. We comprehensively compare Single-Codec with SOTA multi-codebook codecs, including EnCodec and TiCodec, by conducting both objective and subjective tests in analysis-synthesis and TTS. The results show that Single-Codec with the lower bandwidth has better speech reconstruction quality, and significantly improves intelligibility, naturalness, and speaker similarity of synthesized speech in zero-shot LLM-TTS. Audio samples are available at \linebreak \url{https://kkksuper.github.io/Single-Codec}.

\section{Method}

\subsection{Architecture of Single-Codec}
\begin{figure}[t]
  \centering
  \includegraphics[width=8cm]{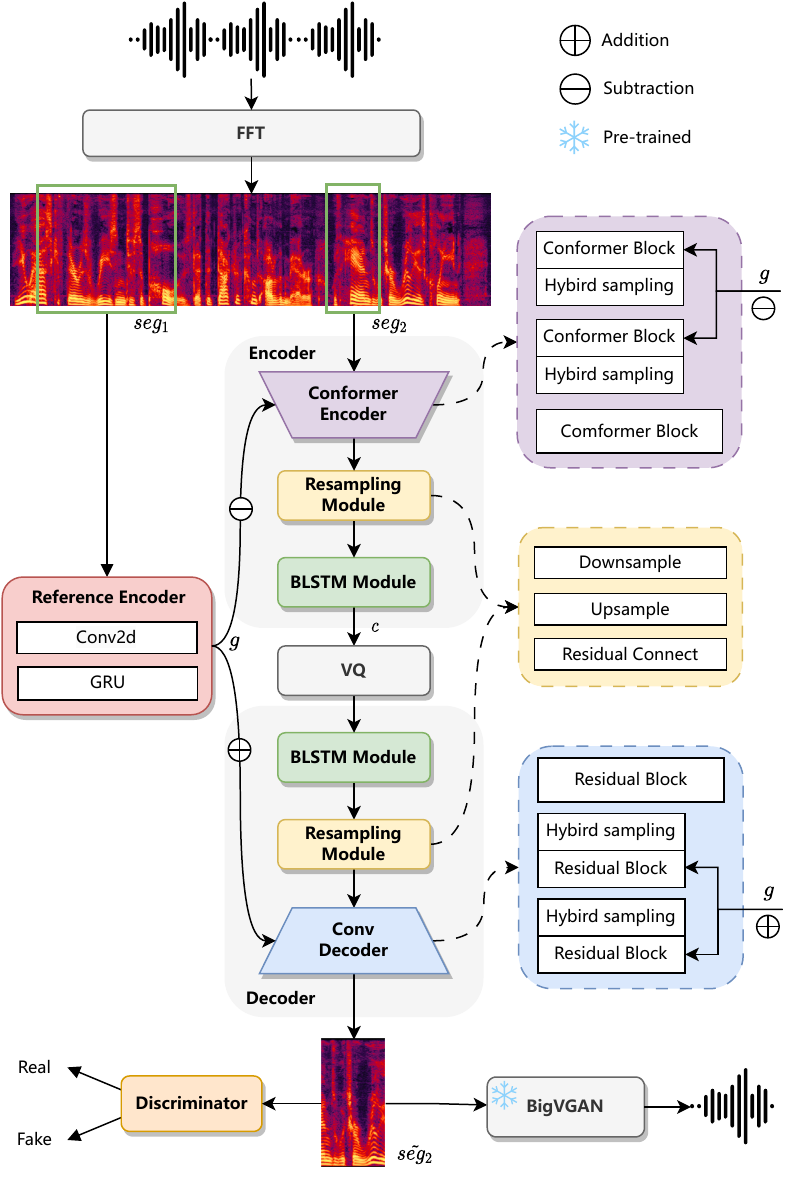}
  \caption{The architecture of Single-Codec.}
  \label{fig:overview}
\end{figure}

The architecture of Single-Codec is shown in Figure~\ref{fig:overview}. It is built on Vector Quantised-Variational AutoEncoder (VQVAE)~\cite{Oord2017NeuralDR} with Mel Spectrogram input and reconstruction, similar to Tortoise TTS~\cite{Betker2023BetterSS}. We adopt a Conformer-based encoder to encode a Mel Spectrogram segment $seg_{2}$ into a latent content representation $c$, which is then passed to the Vector Quantizer (VQ) for vector quantization. The convolution-based decoder reconstructs the Mel Spectrogram $\tilde{seg_{2}}$ from the quantized content representation $c$.  Additionally, we apply a discriminator to improve generation quality~\cite{DBLP:conf/cvpr/EsserRO21}.  Finally, we use a neural vocoder BigVGAN~\cite{Lee2022BigVGANAU} to reconstruct waveform from codec output, i.e., Mel Spectrogram. 

To achieve a high-quality single-codebook codec, we improve the codec architecture with four modules. Specifically, we add a reference encoder to decouple time-invariant information in speech from a Mel Spectrogram segment $seg_{1}$, yielding a global representation $g$. A hybrid sampling module is adopted to alleviate sampling loss. Moreover, we introduce a BLSTM~\cite{Schuster1997BidirectionalRN} module and resampling module in both codec encoder and decoder to enhance contextual information and phonetics-relevant information, respectively.



\subsection{Reference Encoder} 


Speech contains multiple aspects of information, such as time-variant content, time-invariant timbre, and acoustic environment. Multiple codebook in codec makes it easy to encode these various information. However, for a single-codebook codec, it is challenging to compress all information into a limited number of discrete units. To solve this problem, we decouple global information (such as timbre and acoustic environment) that is almost invariable in all frames of the utterance and discretize speech content into code.

We introduce a reference encoder to derive global representation $g$ that is mainly related to timbre and acoustic environment. The input of the reference encoder is a segment $seg_{1}$ randomly selected from the input utterance. We set the length of the segment $seg_{1}$ for reference input to 600 frames while the input segment $seg_{2}$ for codec encoder to 200 frames, where the short segment $seg_{2}$ can reduce the amount of calculation and memory overhead, while the longer reference segment $seg_{1}$ can help to obtain more robust global features. The output $g$ of the reference encoder is fed to the codec encoder and decoder after passing through different linear layers, where it subtracts with output of the encoder blocks and adds to the input of the decoder blocks.

\subsection{BLSTM Module}

Codecs are generally trained on large-scale speech data to ensure good generalization. The diversity of speech content creates challenges for single-codebook codecs with appropriate sizes. Unlike EnCodec~\cite{Defossez2022HighFN}, which introduces the sequence modelling with LSTM~\cite{Hochreiter1997LongSM} and finds that it can improve the Scale-Invariant Signal-to-Noise Ration (SI-SNR)~\cite{DBLP:journals/taslp/VincentGF06}, we add BLSTM modules before and after the quantizer to enhance contextual information. We found this can improve the efficiency of speech content modelling and make it easier to form stable clustering centers. 

\subsection{Hybrid Sampling Module}

Neural codecs usually employ a sampling module to reduce the sequence length of the discrete representation. Currently, the up-sampling and down-sampling operations in codecs are usually implemented by convolution, transposed convolution, or pooling and repeat. The sampling process inevitably produces sampling loss, resulting in reduced encoding and decoding capabilities. Inspired by  MR-HuBERT~\cite{Shi2023MultiresolutionHM}, we introduce an improved hybrid sampling module that uses both convolution and pooling to achieve downsampling and transposed convolution and replication to achieve upsampling. The combination of different sampling methods can alleviate sampling distortion.

\subsection{Resampling Module}


The main goal of a single-codebook speech codec is to extract short-term invariant speech units from acoustic representations. The diversity of acoustic representations brings challenges to the learning of codebook vectors. To solve this problem, we propose a novel resampling module, which first downsamples the input feature for local modelling and then residual connect after upsampling. This bottlenecking operation along the time axis encourages the encoder to extract more phonetics-relevant information with lower short-time variance from the acoustic sequence.

\section{Experiments}

\begin{table*}[]
    \centering
    \caption{Objective metrics scores of various codecs, where the bold numbers highlight best results in 304bps.}
    \label{tab:codec_metrics} 
\begin{tabular}{llcccccc}
\toprule
Experiment                           & Model                 & \multicolumn{1}{l}{Bandwidth (bps)} & \multicolumn{1}{l}{STOI$\uparrow$} & \multicolumn{1}{l}{PESQ$\uparrow$} & \multicolumn{1}{l}{MCD$\downarrow$} & \multicolumn{1}{l}{UTMOS$\uparrow$} & \multicolumn{1}{l}{SPK$\uparrow$} \\ \midrule
\multirow{7}{*}{Ablation}    & VQVAE                 & \multirow{7}{*}{304}                & 0.805                              & 1.591                              & 4.489                               & 2.847                               & 0.727                             \\
                                     & Ref-short             &                                     & 0.827                              & 1.797                              & 4.184                               & 2.884                               & 0.790                             \\
                                     & Ref-long              &                                     & 0.833                              & 1.819                              & 4.122                               & 2.983                               & 0.809                             \\
                                     & Ref-BLSTM             &                                     & 0.837                              & 1.879                              & 4.054                               & 2.933                               & 0.814                             \\
                                     & Ref-HybSam            &                                     & 0.837                              & 1.899                              & 4.077                               & 2.961                               & 0.800                             \\
                                     & Ref-BLSTM-HybSam      &                                     & 0.841                              & 1.912                              & 4.042                               & \textbf{3.081}                      & 0.809                             \\
                                     & Ref-BLSTM-HybSam-Conf &                                     & 0.839                              & 1.901                              & 4.091                               & 3.029                               & 0.811                             \\ \midrule
\multirow{3}{*}{Comparison} & EnCodec-1VQ           & 750                                 & 0.765                              & 1.319                              & 4.515                               & 2.521                               & 0.599                             \\
                                     & TiCodec-1VQ           & 750                                 & 0.808                              & 1.694                              & 4.486                               & 2.919                               & 0.684                             \\
                                     & TiCodec-2VQ           & 1500                                & 0.866                              & 2.140                              & 3.865                               & 3.079                               & 0.762                             \\ \midrule
                                     & Single-Codec          & 304                                 & \textbf{0.842}                     & \textbf{1.933}                     & \textbf{4.017}                      & 3.031                               & \textbf{0.817}                    \\ \bottomrule
\end{tabular} \vspace{-1.2em}
\end{table*}
\vspace{-0.5em}
\subsection{Dataset}
\vspace{-0.5em}
We train speech codecs and a zero-shot TTS system, VALL-E~\cite{Wang2023NeuralCL}, using five open-source datasets, including LibriTTS\cite{Zen2019LibriTTSAC}, Hi-Fi TTS\cite{Bakhturina2021HiFiME}, VCTK\cite{Liu2019CrosslingualMT}, AISHELL-1\cite{Bu2017AISHELL1AO}, and AISHELL-3\cite{Shi2020AISHELL3AM}. A total of 1165.3 hours of English and Chinese speech is used.





\vspace{-0.8em}
\subsection{Comparison Models}
\vspace{-0.2em}
We adopt EnCodec~\cite{Defossez2022HighFN} with one codebook (EnCodec-1VQ) and TiCodec~\cite{Ren2023FewertokenNS} with one codebook (TiCodec-1VQ) and two codebooks (TiCodec-2VQ) as the baselines to compare with our proposed Single-Codec. For VALL-E, we use EnCodec with one, four, and eight codebooks, representing EnCodec-1VQ, EnCodec-4VQ, and EnCodec-8VQ, TiCodec with one codebook (TiCodec-1VQ) as the baselines to evaluate the performance of codecs on speech synthesis.

To verify the effectiveness of our designed modules in Single-Codec, we conduct ablation studies on the following models.
\begin{itemize}

    \item [$\bullet$] \textbf{VQVAE}: A basic VQVAE codec with a discriminator for perceptual loss, the structure and configuration of the VQVAE codec is similar to that in Tortoise TTS~\cite{Betker2023BetterSS}.

    \item [$\bullet$] \textbf{Ref-short}: VQVAE with a reference encoder that consumes a short segment with 200 frames as input.

    \item [$\bullet$] \textbf{Ref-long}: VQVAE with a reference encoder that consumes a long segment with 600 frames as input.  

    \item [$\bullet$] \textbf{Ref-BLSTM}: Ref-long with the BLSTM module to verify the effectiveness of the BLSTM module. 

    \item [$\bullet$] \textbf{Ref-HybSam}: Ref-long with the hybrid sampling module to verify the effectiveness of the hybrid sampling module.

    \item [$\bullet$] \textbf{Ref-BLSTM-HybSam}: Ref-long with the BLSTM and hybrid sampling modules to verify the effectiveness of the combination of BLSTM and hybrid sampling modules. 
   
    \item [$\bullet$] \textbf{Ref-BLSTM-HybSam-Conf}: Ref-BLSTM-HybSam with the Conformer-based encoder, excluding the resampling module. 
    
\end{itemize}

\vspace{-0.8em}
\subsection{Model Parameters and Training Details}
\vspace{-0.2em}
The audio sample rate is 24khz, and the hop length and window length of the Mel Spectrogram are 256 and 1024, respectively. The downsample rate is 4, resulting in a total downsampling of 1024 times (about 23 discrete tokens per second). The codebook size is 8192. The model size of the codec is 256. The sizes of the intermediate hidden states in convolution blocks are 256, 512, and 1024, while the hidden size of the Conformer~\cite{DBLP:conf/interspeech/GulatiQCPZYHWZW20} block is 1024. The reference encoder consists of 6 layers of 2D convolution with a kernel size of 3 and a GRU~\cite{Chung2014EmpiricalEO} layer. The residual block~\cite{DBLP:conf/cvpr/HeZRS16} consists of two residual units. Each residual unit includes 2 one-dimensional convolutions with kernel sizes of 3 and 1 respectively. Discriminator consists of 4 layers of 2D convolution with a kernel size of 5 and 2 layers of 2D convolution with a kernel size of 3. The BLSTM module contains two LSTM layers with a hidden size of 128.

During training, we conduct 300k iterations using a batch size of 1024 on a single V100 GPU for Single-Codec. The baseline model Encodec utilizes with the code reimplemented in HifiCodec\footnote{https://github.com/yangdongchao/AcademiCodec}~\cite{Yang2023HiFiCodecGV}, and is trained for 25 epochs. TiCodec\footnote{https://github.com/y-ren16/TiCodec} is trained for 300k steps with a batch size of 40 on two V100 GPUs. For TTS, we employ VALL-E, reimplemented in Amphion\footnote{https://github.com/open-mmlab/Amphion}~\cite{Zhang2023AmphionAO}, with dynamic batch sizing and a maximum token limit of 4000 per batch. The single-codebook codec only utilizes the AR stage, while the multiple-codebook codec trains both the AR and NAR stages simultaneously. Eight A800 GPUs and 70 epochs are employed for training VALL-E.

\vspace{-0.8em}
\subsection{Ablation Studies}
\vspace{-0.2em}

We calculate STOI~\cite{Taal2010ASO}, PESQ~\cite{recommendation2001perceptual}, Mel cepstral distortion (MCD)~\cite{Kubichek1993MelcepstralDM}, UTMOS\footnote{https://github.com/tarepan/SpeechMOS}~\cite{Saeki2022UTMOSUS} and speaker cosine similarity (SPK)\footnote{We adopt
WeSpeaker~\cite{Wang2022WespeakerAR} to extract speaker embedding for cosine similarity compution} to objectively evaluate the quality of speech reconstruction. The test set is composed of 100 randomly selected sentences from unseen speakers. The objective result is shown in Table~\ref{tab:codec_metrics}. 

Compared with VQVAE, either Ref-short or Ref-long obtains better performance on all metrics.
It indicates that it is effective to decouple global information from speech for the single-codebook codec. Moreover, Ref-long outperforms Ref-short in both reconstruction and speaker similarity, suggesting that longer reference segments help capture more accurate time-invariant information and enhance content modelling. Ref-BLSTM, Ref-HybSam, and Ref-BLSTM-HybSam get higher reconstruction quality, showing the effectiveness of the BLSTM and hybrid sampling modules. Moreover, Ref-BLSTM-HybSam-Con yields on-pair performance with Ref-BLSTM-HybSam but gets further improvement after adding the resampling module, i.e., our proposed Single-Code, achieving the best results.




\begin{figure}[t]
  \centering
  \includegraphics[width=7.4cm]{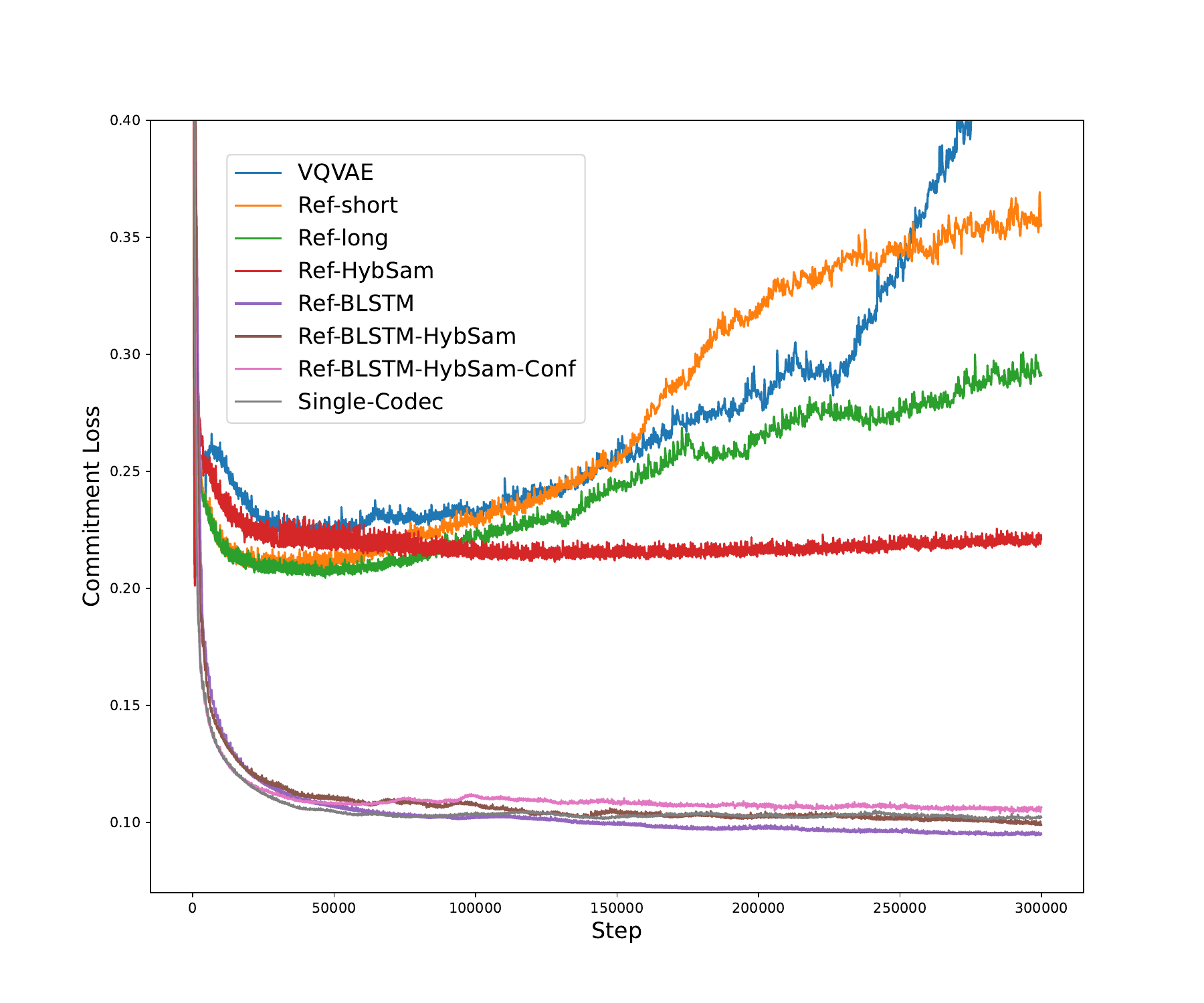}
  \vspace{-1.0em}
  \caption{The commitment loss of different codec while training.}
  \label{fig:commitloss}
  \vspace{-2.0em}
\end{figure}

\vspace{-0.8em}
\subsection{Commitment Loss Analysis}
\vspace{-0.2em}

We further analyze the commitment loss in the training to explore the impact of different designed modules on the single-codebook codec. Commitment loss is the difference between representations before and after quantization. The degree of convergence of the commitment loss can reflect the relationship between the encoder output and the cluster center in the codebook. As shown in  Figure~\ref{fig:commitloss}, the commitment loss of VQVAE tends to diverge after model training, indicating that the entanglement of time-invariant global information and time-variant content information hinders forming a limited variety of content-related speech units. After considering time-invariant decoupled modelling, the loss of Ref-short increases slowly, indicating the effectiveness of global information disentanglement for speech unit learning. Ref-long further verifies this result, illustrating the effectiveness of a longer reference segment. The loss curve of Ref-HybSam is flat, indicating that the hybrid sampling module effectively improves codec performance. Moreover, the losses of the models with context modelling via the BLSTM module are all converged. It demonstrates that the models have learned stable phonetic units before quantization, indicating the effectiveness of context modelling in codecs. 

Furthermore, considering the results presented in Table~\ref{tab:codec_metrics}, we observe that the commitment loss is not strictly inversely related to reconstruction quality. However, the convergence status of the commitment loss (divergence, flat, convergence) is indeed associated with reconstruction quality. Specifically, the converged codec surpasses the codec which is not converged. This result further highlights the significance of achieving a stable clustering center in the single codebook codec, which directly impacts the overall reconstruction quality.

\vspace{-0.8em}
\subsection{Speech Reconstruction Evaluation}
\vspace{-0.2em}

We compare the performance in speech reconstruction of the proposed Single-Codec with other codecs. The results, as presented in Table~\ref{tab:codec_metrics}, demonstrate that despite lower bandwidth, the proposed Single-Codec surpasses other codecs with 1 codebook and is on par with the TiCodec with 2 codebooks in terms of reconstruction quality and speaker similarity. VQVAE performs better than EnCodec with 1 codebook, demonstrating the high quantization efficiency of codecs operated on Mel Spectrogram. Compared to TiCodec, which also quantized the decoupled time-invariant information, Single-Codec achieves higher speaker similarity and reconstruction quality, indicating the effectiveness of continuous time-invariant representations and longer reference length.

\vspace{-0.8em}
\subsection{Zero-shot TTS Evaluation}
\vspace{-0.2em}

To evaluate the performance of codecs applied in speech synthesis tasks, we train VALL-E~\cite{Wang2023NeuralCL} using discrete tokens extracted from EnCodec in the number of 1,4,8 codebooks, TiCodec in 1 codebook, and Single-Codec. We conduct naturalness Mean Opinion Score (N-MOS) and speaker similarity MOS (S-MOS) for subjective evaluation of the synthesized speech. The test set consists of 30 sentences, including Chinese and English speech. The 20 listeners who are Chinese Mandarin native speakers and familiar to English are invited to participate in each MOS test. Meanwhile, we calculate the word error rate (WER) using an ASR model\footnote{https://huggingface.co/openai/whisper-large}~\cite{Radford2022RobustSR} to measure speech intelligibility. We also use WeSpeaker~\cite{Wang2022WespeakerAR} to extract speaker embedding to calculate the speaker embedding cosine similarity.

Table~\ref{tab:tts} shows the subjective and objective results. Single-Codec outperforms other models in terms of naturalness and speaker similarity. In single-codebook scenes, TiCodec-1VQ and Single-Codec are significantly better than other codec models in speaker similarity, naturalness, and stability. This is because decoupling global information makes the frame-level codebook pay more attention to content modelling and enables more global information transmission. Meanwhile, Single-Codec performs better than Ticodec, indicating the effectiveness of continuous global representation and additional content modelling. In addition, Single-Codec exceeds multiple-codebook codecs regarding speaker similarity and naturalness while WER is slightly higher than Encodec-8VQ. This is mainly because the higher bandwidth brings higher-resolution speech unit perception.

\begin{table}[]
    \centering
    \caption{The objective and subjective metrics scores of VALL-E.}
    \label{tab:tts}
\scalebox{0.85}{
\begin{tabular}{lcccc}
\toprule
\multicolumn{1}{c}{\multirow{2}{*}{Codec Model}} & \multicolumn{2}{c}{Subjective}    & \multicolumn{2}{c}{Objective}                            \\
\multicolumn{1}{c}{}                       & N-MOS$\uparrow$ & S-MOS$\uparrow$ & WER$\downarrow$ & \multicolumn{1}{l}{S-Cosine$\uparrow$} \\ \midrule
VQVAE                                        & 2.81 $\pm$ 0.06            & 3.24 $\pm$ 0.07           & 30.9            & 0.649                                  \\
EnCodec-1VQ                                & 3.36 $\pm$ 0.06            & 3.68 $\pm$ 0.08            & 28.7            & 0.669                                  \\
EnCodec-4VQ                                & 3.70 $\pm$ 0.10            & 3.86 $\pm$ 0.08            & 15.9            & 0.725                                  \\
EnCodec-8VQ                                & 3.77 $\pm$ 0.12            & 4.02 $\pm$ 0.11            & \textbf{11.4}   & 0.738                                  \\
TiCodec-1VQ                                & 3.68 $\pm$ 0.08            & 3.97 $\pm$ 0.07            & 19.4            & 0.683                                  \\ \midrule
Single-Codec                               & \textbf{4.02 $\pm$ 0.07}   & \textbf{4.13 $\pm$ 0.09}   & 13.2            & \textbf{0.792}                         \\ \bottomrule 
\end{tabular}
}    \vspace{-1.0em}

\end{table}


\section{Conclusion}


In this paper, we propose Single-Codec, the first
single-codebook codec dedicatedly designed for LLM-based
speech generation. Single-Codec employs a disentangled VQ-VAE on Mel Spectrograms to decouple speech into time-invariant global embedding and one phonetically-rich discrete sequence quantized by one codebook. Furthermore, the encoder is enhanced with a BLSTM module for contextual modelling, a hybrid sampling module to alleviate distortion from upsampling and downsampling, and a resampling module to encourage discrete units to carry more phonetic-relevant information with lower short-time variance. In experiments, compared with multi-codebook codecs, e.g. EnCodec and TiCodec, Single-Codec demonstrates higher speech reconstruction quality with lower bandwidth of only 304bps, and enables a higher-quality LLM-TTS with better naturalness and intelligibility. In the future, we will focus on developing a more efficient single-codebook codec for speech reconstruction and speech synthesis.

\bibliographystyle{IEEEtran}
\bibliography{mybib}

\end{document}